\documentclass[twocolumn,prd,showpacs,superscriptaddress]{revtex4}
\usepackage{amssymb}
\usepackage{amsmath}
\usepackage{graphicx}
\usepackage{subfigure}
\usepackage{natbib}
\usepackage{epsfig}
\usepackage{amsfonts}
\usepackage{mathrsfs}
\usepackage{CJK}
\usepackage[toc,page,title,titletoc,header]{appendix}

\begin{document}

\title{Quantum speedup dynamics process in Schwarzschild space-time}

\author{Kai Xu}
 \affiliation{Key Laboratory of Micro-Nano Measurement-Manipulation and Physics (Ministry of
Education), School of Physics, Beihang University,
Xueyuan Road No. 37, Beijing 100191, China}

\author{Han-Jie Zhu}
 \affiliation{Beijing National Laboratory for Condensed Matter Physics, Institute of Physics, Chinese Academy of Sciences, Beijing 100190, China}

\author{Guo-Feng Zhang}
\email{gf1978zhang@buaa.edu.cn}
 \affiliation{Key Laboratory of Micro-Nano Measurement-Manipulation and Physics (Ministry of
Education), School of Physics, Beihang University,
Xueyuan Road No. 37, Beijing 100191, China}

\author{Jie-Ci Wang}
 \affiliation{Department of Physics and Key Laboratory of Low Dimensional Quantum Structures
and Quantum Control of Ministry of Education, Hunan Normal University, Changsha,
Hunan 410081, China}

\author{Wu-Ming Liu}
 \affiliation{Beijing National Laboratory for Condensed Matter Physics, Institute of Physics, Chinese Academy of Sciences, Beijing 100190, China}
 \affiliation{School of Physical Sciences, University of Chinese Academy of Sciences, Beijing 100190, China}
 \affiliation{Songshan Lake Materials Laboratory, Dongguan, Guangdong 523808, China}

\date{\today}
\begin{abstract}
Quantum speed limit time (QSLT) can be used to characterize the intrinsic minimal time interval for a quantum system evolving
from an initial state to a target state. We investigate the QSLT of the open system in Schwarzschild space-time.
We show that, in some typical noisy channels, the Hawking effect can be beneficial to the evolution of the system.
For an initial entangled state, the evolution speed of the system can be enhanced in the depolarizing, bit flip, and bit-phase flip channels 
as the Hawking temperature increases,  which are in sharp contrast to the phase flip channel. 
Moreover, the optimal initial entanglement exists in other noise channels except the phase flip channel, 
which minimizes the QSLT of the system and thus leads to the maximum evolution speed of the system.
\end{abstract}
\pacs {03.65.Yz, 03.67.Lx, 42.50.-p}

\maketitle 
     
\section{\textbf{{Introduction}}}

The quantum information theory under the relativistic framework  
has attracted much attention due to its theoretical importance and practical application \cite{PeresA,PanC123}.
It is necessary to understand quantum phenomena in the framework of relativity theory
because the realistic quantum systems are essentially noninertial.
Experiments have shown \cite{Chou123} that when the height of atomic clocks changes to 33 cm, 
the gravity frequency shift effect has a significant effect on the accuracy of atomic clocks.
Furthermore, many theoretical researchers 
have considered the dynamical behaviors of the system in the context 
of relativity \cite{Mann123q,q2,Bruschiq,Yaoq,Bruschi2q,Ahmadiq,Daiq}.
In particular, the Hawking effect, i.e., the particle creation process that gives rise to 
a thermal spectrum of radiation outgoing from a black hole, has a significant effect on the 
quantum entanglement \cite{JingJ,Song}, quantum discord  \cite{Wangq} 
and some other information quantities  \cite{Wu321}.

Recently, much attention has been given to the study of the quantum speed limit time (QSLT)  \cite{Tamm,Margolus,Mirkin,ZhangX,Lidar,Taddei,Egusquiza,Pires,Pollock,Sun2,Wang0,
Liuq,Hanq,Wu1} of the 
system in the relativistic context \cite{Haselia}.
A lower bound on the time required for
a quantum system to evolve between two given states, which is defined as the QSLT,
is a key factor in characterizing the maximal speed of evolution of quantum systems. 
It is widely used in many fields such as quantum metrology \cite{Maccone1,Demkowicz},
quantum communication \cite{Nielsen,Bekenstein}, nonequilibrium thermodynamics \cite{Deffner3},
and quantum optimal control algorithms \cite{Hegerfeldt,Hegerfeldt1,Avinadav}.
Researchers have considered the QSLT of the system under the relativistic scenario. 
Haseli $et$ $al$. \cite{Haselia} studied the QSLT of a single qubit system in Schwarzschild space-time and showed a fixed distance $r_{0}$ 
outside the event horizon the influence on the QSLT of the system. It is worth noting that the investigation of
the QSLT in Schwarzschild space-time focused on the initial single-qubit state (i.e., the initial unentangled state).
However, quantum entanglement is both the central concept and the
most satisfying resource for all kinds of quantum information
processing tasks \cite{Boschi,Bouwmeester}, such as superdense coding, quantum teleportation,
error-correcting codes, and quantum computation. Since the QSLT depends on the choice of the initial state \cite{Wu123},
it is necessary to consider the influence of the Hawking effect on the QSLT when the system is in the initial entangled state, 
and explore what initial conditions can realize the maximum evolution of the system in Schwarzschild space-time.

To do this, in this paper, we take the initial GHZ-like state as an example to discuss the influence of the Hawking effect on the QSLT of the system.
More specifically, we consider a situation in which Alice, Bob, and Charlie share a generically
tripartite GHZ-like state at the same initial point in flat
Minkowski space-time. At one moment, one of them
freely falls in toward a Schwarzschild black hole and
locates near the event horizon, the others remain at
the asymptotically flat region.
Then in the depolarizing, bit flip, and bit-phase flip channels, we find that the evolution speed of the system 
can be enhanced by increasing the Hawking temperature when the system is initial entangled state.
However, in the case of the initial unentangled state, the capacity for potential speedup of quantum system in a Pauli channel cannot
be increased with the increase of the Hawking temperature.
In addition, in other channels except the phase flip channel,
we show that, by fixing the decoherence parameter of a Pauli channel, 
the optimal initial entanglement of the system can be obtained, so as to minimize the QSLT of the system and thus
to achieve the optimal evolution speed of the system.

This paper is organized as follows. In Sec. II 
we review the vacuum structure
for Dirac fields in the Schwarzschild space-time and introduce the QSLT.
Sec. III discusses the influence of Hawking effect on the dynamical speedup of
the quantum evolution for different Pauli noisy models in three subsection.
In Sec. IV, the relationship between quantum speed limit time and initial entanglement is discussed.
The conclusions drawn from the present study are given in Sec. V.

\section{\textbf{Preliminaries}}

To begin with, we introduce a concise review of vacuum structure 
for Dirac particles in Schwarzschild space-time. The metric for the Schwarzschild space-time can be given as 
$d s^{2}=-\left(1-\frac{2 M}{r}\right) d t^{2}+\left(1-\frac{2 M}{r}\right)^{-1} d r^{2}+r^{2}\left(d \theta^{2}+\sin ^{2} \theta d \varphi^{2}\right)$.
The Dirac 
equation \cite{Brillq} $\left[\gamma^{a} e_{a}^{\mu}\left(\partial_{\mu}+\Gamma_{\mu}\right)\right] \psi=0$
in Schwarzschild space-time can be written as 
$-\gamma_{0}\partial_{t} \psi/\sqrt{1-2 M/r}+\gamma_{1} \sqrt{1-2 M/r}[\partial_{r}+1/r+M/(2 r(r-2 M))] \psi 
+\gamma_{2}\left(\partial_{\theta}+\cot \theta/2\right) \psi/r+\gamma_{3}\partial_{\phi} \psi/(r \sin \theta) =0$,
where the parameter $M$ represents the mass of the black hole, and we consider $G$, $c$, $\hbar$ and $\kappa_B$ as unity 
in this paper for simplicity.

By solving the  Dirac equation in Schwarzschild space-time, we can obtain the positive (fermions) frequency 
outgoing solutions for the outside and inside regions near the event horizon \cite{Jingq123}
$\psi_{k}^{\mathrm{I}{+}}\left(r>r_{+}\right)=\Re e^{-i \omega u},\psi_{k}^{\mathrm{II}{+}}\left(r<r_{+}\right)=\Re e^{i \omega u}$, where $u=t-r^{*}$
and $r^{*}=r+2 M \ln [(r-2 M) /(2 M)]$ represent the tortoise coordinate, and $\Re$ is a 4-component Dirac spinor \cite{Wangqw}.
Then, in above bases, the Dirac field can be expressed as 
$\psi_{\text {out}}=\sum_{\sigma} \int d k\left(a_{k}^{\sigma} \psi_{k}^{\sigma+}+b_{k}^{\sigma+} \psi_{k}^{\sigma-}\right)$, where $\sigma=(I, I I)$,
$I$ and $II$ correspond to the state of exterior and interior region respectively, $a_{k}^{\sigma}$ and $b_{k}^{\sigma+}$
denote the fermion annihilation and antifermion creation operators.

On the other hand, according to the suggestion of Damour$-$Ruffini \cite{Damoarq}, we get a complete basis for positive energy modes after making an analytic continuation
for $\psi_{k}^{\mathrm{I}{+}}$ and $\psi_{k}^{\mathrm{II}{+}}$. Then, by quantizing the Dirac fields in the Schwarzschild and Kruskal modes respectively,
the Bogoliubov transformations \cite{Barnettq} between the creation operator and the annihilation operator in the Schwarzschild and Kruskal coordinates can be obtained.
After a series of calculation, the vacuum and excited states of the Kruskal particle for mode $\mathbf{k}$ can be expressed as \cite{Song}

\begin{equation}\begin{aligned}
|0\rangle_{K}=&\left(e^{-\omega_{k} / T}+1\right)^{-1 / 2}\left|0_{k}\right\rangle_{I}^{+}\left|0_{-k}\right\rangle_{I I}^{-} \\
&+\left(e^{\omega_{k} / T}+1\right)^{-1 / 2}\left|1_{k}\right\rangle_{I}^{+}\left|1_{-k}\right\rangle_{I I}^{-},\\
|1\rangle_{K}=&\left|1_{k}\right\rangle_{I}^{+}\left|0_{-k}\right\rangle_{I I}^{-},
\end{aligned}\end{equation}
where $T=1 / (8 \pi M)$ is the Hawking temperature, $\{|n_{k}\rangle_{I}^{+}\}$ and $\{|n_{-k}\rangle_{I I}^{-}\}$
are the orthogonal bases for the inside and outside regions of the event horizon respectively, and the superscript on the 
kets $\{+,-\}$ indicate the particle and antiparticle vacua. In addition, hereafter, we consider $\{|n_{k}\rangle_{I}^{+}\}$ 
and $\{|n_{-k}\rangle_{\Pi}^{-}\}$ as $\left\{|n\rangle_{I}\right\}$ and $\left\{|n\rangle_{\mathrm{II}}\right\}$ for convenience.

Then in order to investigate the evolution speed of quantum state of open system in Schwarzschild space-time, 
we need to start with the definition of QSLT for an open quantum system. QSLT can effectively define the bound of minimal evolution time
for arbitrary initial states $\rho_{\mathrm{s}}(0)$ to the corresponding target evolution state $\rho_{\mathrm{s}}(\tau)$, and facilitate to analyze the
maximal evolutional speed of open quantum system. Recently, Campaioli $et$ $al$. choose the Euclidean distance
$D(\rho_{\mathrm{s}}(0),\rho_{\mathrm{s}}(\tau)) =\left\|\rho_{\mathrm{s}}(0)-\rho_{\mathrm{s}}(\tau)
\right\|_{\mathrm{hs}} $ to acquire a robust and feasible QSLT \cite{Campaioli1}

\begin{equation}
\tau_{\mathrm{QSL}}=\frac{\left\|\rho_{\mathrm{s}}(0)-\rho_{\mathrm{s}}(\tau)
\right\|_{\mathrm{hs}}}{\overline{\left\|\dot{\rho}_{\mathrm{s}}(t)\right\|}_{\mathrm{hs}}},
\end{equation}
where $\overline{\left\|\dot{\rho}_{\mathrm{s}}(t)\right\|}_{\mathrm{hs}}=(1/\tau)
\int_{0}^{\tau} \mathrm{d} t\left\|\dot{\rho}_{\mathrm{s}}(t)\right\|_{\mathrm{hs}}$,
$\|A\|_{\mathrm{hs}}=\sqrt{\sum_{i} r_{i}^{2}}$ and $r_{i}$ is the singular values of $A$.
When the ratio between the QSLT and the actual evolution time
equals unity, i.e., $\tau_{QSL}/\tau=1$, the quantum system evolution
is already along the fastest path and possesses no potential
capacity for further quantum speedup. While $0<\tau_{QSL}/\tau<1$, the
speedup evolution of the quantum system may occur, and the
smaller $\tau_{QSL}/\tau$, the greater the capacity for potential speedup
will be.

\section{\textbf{QSLT in SCHWARZSCHILD SPACE-TIME}}

In this section, we study the impact of Hawking effect on the QSLT of the system.
We assume a situation that Alice, Bob and Charlie share a generically entangled state at the 
same initial point in flat Minkowski space$–$time. Alice has a detector which 
only detects mode $|n\rangle_{A}$, and Bob and Charlie has his detector sensitive only for mode
$|n\rangle_{B}$ and $|n\rangle_{C}$, respectively. At one moment, Alice and Bob remain at the asymptotically
flat region, but Charlie freely fall in toward a Schwarzschild black hole as well as 
locate near the event horizon. They share a GHZ-like state 
$|\phi\rangle=\alpha|000\rangle+\sqrt{1-\alpha^{2}}|111\rangle$, where $\alpha\in[0,1]$.
We then utilize Eq. (1) to rewrite $|\phi\rangle$ in terms of Minkowski mode for Alice, Bob 
and black hole mode for Charlie. Due to the interior region being causally disconnected
from the exterior region of the black hole, we can derive the physically accessible 
density matrix $\rho_{ABC_{I}}$ by tracing
over the state of the interior region,

\begin{equation}\begin{aligned}
\rho_{A B C_{I}}=& \alpha^{2} m^{2}|000\rangle\langle 000|+\alpha^{2} n^{2}| 001\rangle\langle 001|+(1-\alpha^{2})\\
&|111\rangle\langle 111| +\alpha m \sqrt{1-\alpha^{2}}(|000\rangle\langle 111|+| 111\rangle\\
&\langle 000|),
\end{aligned}\end{equation}
where $m=\left(e^{-\omega / T}+1\right)^{-1 / 2}$ and $n=\left(e^{\omega / T}+1\right)^{-1 / 2}$.

Based on the fact that the system inevitably interacts with the environment, 
we consider Alice and Bob's state in a Pauli noisy environment. The noisy environment can be 
described by a quantum channel $\mathcal{E}$. If $\mathcal{E}$ acts on the system,
the output state will be given by 

\begin{equation}
\rho_{t}=\sum_{i} p_{i_{1}i_{2}}(\sigma_{i_{1}}\otimes\sigma_{i_{2}}\otimes\sigma_{0}) \rho_{A B C_{I}}(\sigma_{i_{1}} \otimes\sigma_{i_{2}}\otimes \sigma_{0}),
\end{equation}
where $p_{i_{1}i_{2}}=p_{i_{1}}p_{i_{2}}$ represents the joint probability and $\sigma_{1,2,3}$ ($\sigma_{0}$) are Pauli (identity) operators. 
For convenience, we focus on bit flip channel (BFC), bit-phase flip channel (BPFC), phase flip channel (PFC), and depolarizing channel (DPC), which belong to the 
class of Pauli channels of Eq. (4). The parameter $p_{i}$ for the former three channels 
can be represented by $p_{0}=p$ and $p_{i}=1-p$ (with $i=1$ for BFC, $i=2$ for BPFC, $i=3$ for PFC),
while they are $p_{0}=p$ and $p_{1,2,3}=(1-p)/3$ for DPC. Here $p$ is decoherence parameter. By combining these expressions with equation (4),
one can obtain the output state in the standard computational basis $\{|1\rangle=|111\rangle,|2\rangle=|110\rangle,|3\rangle=|101\rangle,|4\rangle=|100\rangle
,|5\rangle=|011\rangle,|6\rangle=|010\rangle,|7\rangle=|001\rangle,|8\rangle=|000\rangle\}$.
 
In the following, to fully understand the influence of Hawking effect on the evolution speed of the system, we  take the initial 
unentangled state and entangled state of the system in Pauli channels as examples to study the speedup evolution of quantum state.

\subsection{\textbf{QSLT in the DPC}}

\begin{figure}[tbh]
\includegraphics*[bb=37 119 291 368,width=9cm, clip]{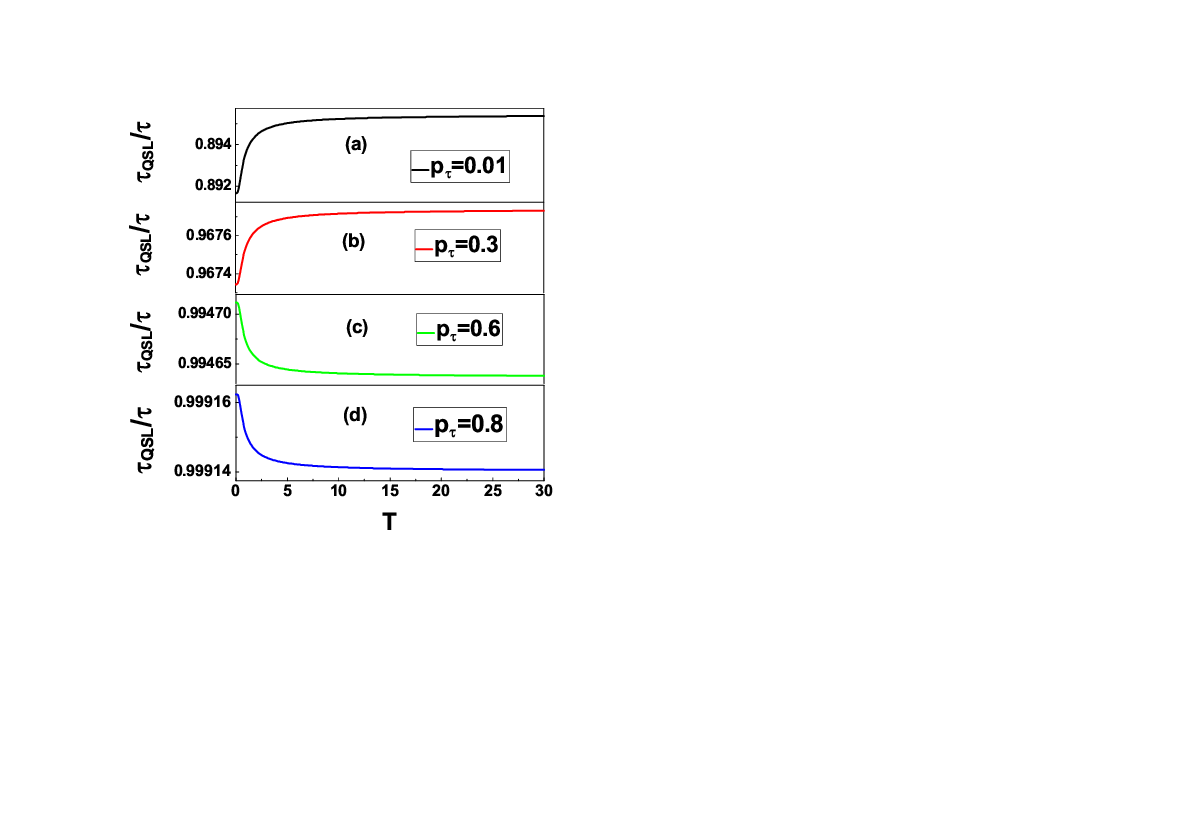}
\caption{(Color online) QSLT for a given three-qubit state,
quantified by $\tau_{QSL}/\tau$ as a function of Hawking temperature
 $T$ with $\omega=1, \alpha=1/4$, when Alice and Bob's state
are in the DPC.}
\end{figure}

We first consider that Alice and Bob's state in the DPC. For the input
state $\rho_{A B C_{I}}$, the nonzero elements of the evolved density matrix for the system
can be obtained as

\begin{equation}\begin{aligned}
\rho_{11}=&\frac{4(-1+p)^{2} \alpha^{2}+\left(1+e^{\frac{\omega}{T}}\right)(\beta+2 p \beta)^{2}}{9\left(1+e^{\frac{\omega}{T}}\right)},\\
\rho_{22}=&\frac{4 e^\frac{\omega}{T}(-1+p)^{2} \alpha^{2}}{9\left(1+e^{\frac{\omega}{T}}\right)},\\
\rho_{33}=&\rho_{55}=-\frac{2(-1+p)(1+2 p)(\alpha^{2}+(1+e^{\frac{\omega}{T}}) \beta^{2})}{9(1+e^{\frac{\omega}{T}})},\\
\rho_{44}=&\rho_{66}=-\frac{2 e^{\frac{\omega}{T}}(-1+p)(1+2 p) \alpha^{2}}{9\left(1+e^{\frac{\omega}{T}}\right)},\\
\rho_{77}=&\frac{(\alpha+2 p \alpha)^{2}+4\left(1+e^{\frac{\omega}{T}}\right)(-1+p)^{2} \beta^{2}}{9\left(1+e^{\frac{\omega}{T}}\right)},\\
\rho_{88}=&\frac{e^\frac{\omega}{T}(\alpha+2 p \alpha)^{2}}{9\left(1+e^{\frac{\omega}{T}}\right)},\\
\rho_{18}=&\rho_{18}=\frac{(1-4 p)^{2} \alpha \beta}{9 \sqrt{1+e^{-\frac{\omega}{T}}}},
\end{aligned}\end{equation}
where $\beta=\sqrt{1-\alpha^{2}}$.

Based on Eqs. (2) and (5), we can get the QSLT of the system. For the initial unentangled state (i.e., $\alpha=1$ in Eq. (3)), 
the QSLT for system can be simplified as $\tau_{QSL}/\tau=(1-p)\sqrt{11+8p(1+p)}/\int_{p_{\tau}}^{1} \mathrm{d} p \sqrt{11+16p(-1+2p)}$.
We notice that the QSLT of the system is independent of the Hawking temperature $T$. This means that, for initially unentangled state in the DPC, 
the QSLT of the system cannot change with the change of the Hawking temperature.
However, for the initial entangled state (i.e., $\alpha=1/4$) as shown in Fig. 1, the QSLT of the system can 
change monotonically as the Hawking temperature increases. More specifically, when the decoherence parameter $p_{\tau}<p_{\tau_{c}}$ ($p_{\tau_{c}}$ means a certain
critical value of $p_{\tau}$) in Figs. 1(a) and 1(b), QSLT increases monotonically with increasing $T$, which means
that the increase of the Hawking temperature $T$ can inhibit the evolution speed of quantum state. However, 
in the case of $p_{\tau}=0.6,0.8>p_{\tau_{c}}$ in Figs. 1(c) and (d), QSLT decreases
monotonically with the growth of the acceleration parameter $T$. That is to say, for the initial entangled state in the DPC, 
a relatively large decoherence paramenter $p_{\tau}$ is required to realize the purpose that the capacity for potential speedup of quantum system
increases with increasing $T$.

\subsection{\textbf{QSLT in the PFC}}

Next, we consider Alice and Bob's state in the PFC, which describes a decoherencing process
without exchanging energy with the environment.
Based on Eq. (4), we can obtain the evolutional density matrix

\begin{equation}\begin{aligned}
\rho_{t}=& \alpha^{2} m^{2}|000\rangle\langle 000|+\alpha^{2} n^{2}| 001\rangle\langle 001|+(1-\alpha^{2})|111\rangle\\
&\langle 111| +\alpha m \sqrt{1-\alpha^{2}}(1-2p)^{2}(|000\rangle\langle 111|+| 111\rangle \\
&\langle 000|),
\end{aligned}\end{equation}
where $m=\left(e^{-\omega / T}+1\right)^{-1 / 2}$ and $n=\left(e^{\omega / T}+1\right)^{-1 / 2}$.
Then according to Eq. (3) and Eq. (6), $\|\rho_{s}(0)-\rho_{s}(\tau)\|_{\mathrm{hs}}$=$4 \sqrt{2}(1-p_{\tau})p_{\tau}\alpha \sqrt{1-\alpha^{2}}/\sqrt{1+e^{-\omega/ T}}$ and
$\overline{\|\ \dot{\rho}_{s}(t)\|}_{\mathrm{hs}}$=$\int_{p_{\tau}}^{1} \mathrm{d} p  4 \sqrt{2}|1-2p| \alpha \sqrt{1-\alpha^{2}}/\sqrt{1+e^{-\omega/ T}}$ can be obtained.
When we consider the initial state of the system as unentangled state (i.e., $\alpha=1$), $\|\rho_{s}(0)-\rho_{s}(\tau)\|_{\mathrm{hs}}$=0 can be got.
This means that the system stays in its original unentangled state in the PFC. This phenomenon is caused by the fact that PFC is a noisy process of pure quantum mechanical property.
Differently, for the initial entangled state, QSLT can be expressed as

\begin{equation}\frac{\tau_{\mathrm{QSL}}}{\tau}=\left\{\begin{array}{ll}
1, & p_{\tau} \geqslant \frac{1}{2} \\
\frac{2p_{\tau}\left(1-p_{\tau}\right)}{1-2p_{\tau}\left(1-p_{\tau}\right)}, & p_{\tau}<\frac{1}{2}.
\end{array}\right.\end{equation}
We find that QSLT of the system is independent of Hawking temperature for the initial entangled states in the PFC. 
Hence, in the PFC, the evolution speed of the system would not be enhanced with the increase of Hawking temperature, 
whether the system is initially in entangled or unentangled state.

\subsection{\textbf{QSLT in the BFC and BPFC}}

\begin{figure}[tbh]
\includegraphics*[bb=64 175 296 418,width=7.5cm, clip]{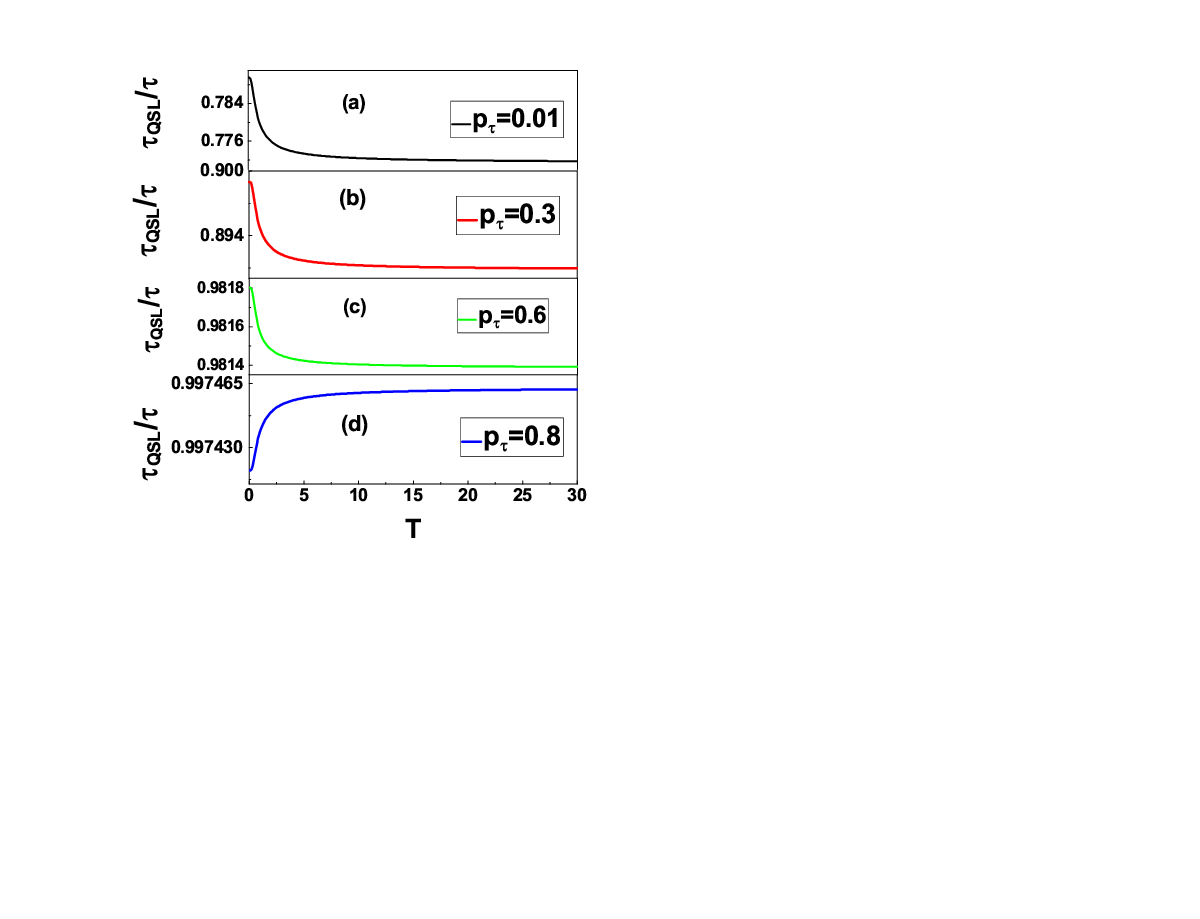}
\caption{(Color online) QSLT for a given three-qubit state,
quantified by $\tau_{QSL}/\tau$ as a function of Hawking temperature
 $T$ with $\omega=1, \alpha=1/4$, when Alice and Bob's state
are in the BFC.}
\end{figure}

Now let us analyze the effect of Hawking temperature $T$ on QSLT by taking Alice and Bob's state in the BFC as an example.
According to Eqs. (3) and (4), the evolved density
matrix of the three-qubit system, whose elements in the standard computational basis
are

\begin{equation}
\begin{aligned} \rho_{11} &=\frac{(-1+p)^{2} \alpha^{2}}{1+e^{\frac{\omega}{T}}}+p^{2} \beta^{2}, \\
\rho_{22} &=\frac{(\alpha-p \alpha)^{2}}{1+e^{-\frac{\omega}{T}}},\\
\rho_{33} &=\rho_{55}=(1-p) p\left(\frac{\alpha^{2}}{1+e^{\frac{\omega}{T}}}+\beta^{2}\right), \\
\rho_{44} &=\rho_{66}=\frac{(1-p) p \alpha^{2}}{1+e^{-\frac{\omega}{T}}}, \\
\rho_{77} &=\frac{p^{2} \alpha^{2}}{1+e^{\frac{\omega}{T}}}+(-1+p)^{2} \beta^{2}, \\
\rho_{88} &=\frac{p^{2} \alpha^{2}}{1+e^{-\frac{\omega}{T}}}, \\
\rho_{18} &=\rho_{81}=\frac{p^{2} \alpha \beta}{\sqrt{1+e^{-\frac{\alpha}{T}}}},\\
\rho_{27} &=\rho_{72}=\frac{(-1+p)^{2} \alpha \beta}{\sqrt{1+e^{-\frac{\omega}{T}}}},\\
\rho_{63} &=\rho_{36}=\rho_{54}=\rho_{45}=-\frac{(-1+p) p \alpha  \beta}{\sqrt{1+e^{-\frac{\omega}{T}}}},
\end{aligned}
\end{equation}
where $\beta=\sqrt{1-\alpha^{2}}$.

Based on Eqs. (2) and (8), we can analyze the effect of the Hawking temperature $T$ on QSLT. For the case of initial unentangled state of the system, the QSLT can be written as 
$\tau_{QSL}/\tau=(1-p)\sqrt{1+2p^{2}}/\int_{p_{\tau}}^{1} \mathrm{d} p \sqrt{3+8p(-1+p)}$, suggesting that the speedup evolution of the system is independent of Hawking temperature. 
As for the case of the initial entangled state, the analysis results we present are drawn in Fig. 2.
We find that, by fixing the decoherence parameter $p_{\tau}<p_{\tau_{2}}$ ( $p_{\tau_{2}}$ means a certain
critical value of $p_{\tau}$) in Figs. 2(a)-(c), the QSLT of the system decreases monotonically as $T$ increases.
This implies that the capacity for potential speedup of system can
be improved by increasing Hawking temperature $T$. 
Differently, in the case of $p_{\tau}>p_{\tau_{2}}$, QSLT increases monotonically with the increase of $T$, illustrating that 
the evolution speed of the system can be weakened with the increase of $T$.
Therefore, for the initial entangled state in BFC, small decoherence parameters can achieve the purpose of accelerating the evolution of the 
system with the increase of $T$, which is in sharp contrast with the system in the DPC.

For the Alice and Bob's state traversing the BPFC, the
density matrix of the output state has a similar form to that for
the BFC. The only difference is that the matrix elements
$\rho_{63}$, $\rho_{36}$, $\rho_{45}$, $\rho_{54}$ are multiplied by a minus. One can then show that
the resulting QSLT have completely the same form to those
for the BFC.

\section{\textbf{THE EFFECT OF the entanglement
ON QSLT in Schwarzschild space-time}} 

In the previous section we have shown that the Hawking temperature may have a favorable effect
on the accelerated evolution of the system which is initially entangled. Then one might wonder
what initial conditions will lead to optimal evolution of the system in Schwarzschild space-time $?$ 
To solve this problem, in this section, we examine the relationship between QSLT and initial entanglement.
The entanglement of the system can be characterized by genuinely multipartite (GM) concurrence \cite{Hashemi Rafsanjani}.
For the initial three-qubit state $\rho_{A B C_{I}}$, the GM concurrence 
$C=2 \alpha \sqrt{1-\alpha^{2}}(1+e^{-\frac{\omega}{T}})^{-1/2}$ can be obtained.
Clearly, we can get the maximum initial entanglement $C_{max}=(1+e^{-\frac{1}{3}})^{-1/2}\approx 0.76$
by fixing $\omega=1$ and $T=3$. In what follows, we would discuss the relationship between QSLT and $C$ 
in the DPC, BFC and BPFC, since the initial entanglement has nothing to do with QSLT in the PFC.

\subsection{\textbf{In the DPC}}

\begin{figure}[tbh]
\includegraphics*[bb=61 141 311 392,width=8cm, clip]{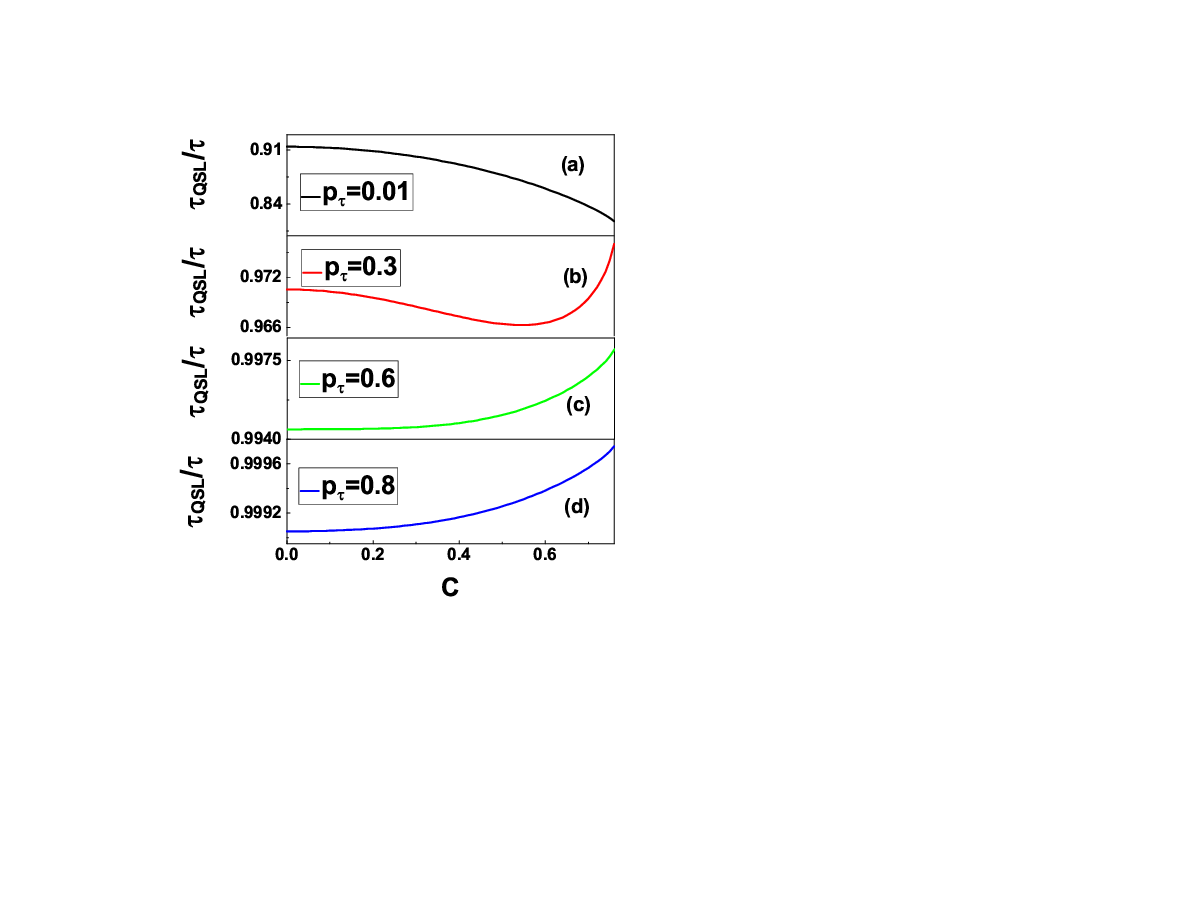}
\caption{(Color online) QSLT for a given three-qubit state,
quantified by $\tau_{QSL}/\tau$ as a function of the initial entanglement
 $C$ with $\omega=1, T=3$, when Alice and Bob's state
are in the DPC.}
\end{figure}

\begin{figure}[tbh]
\includegraphics*[bb=85 124 373 390,width=8cm, clip]{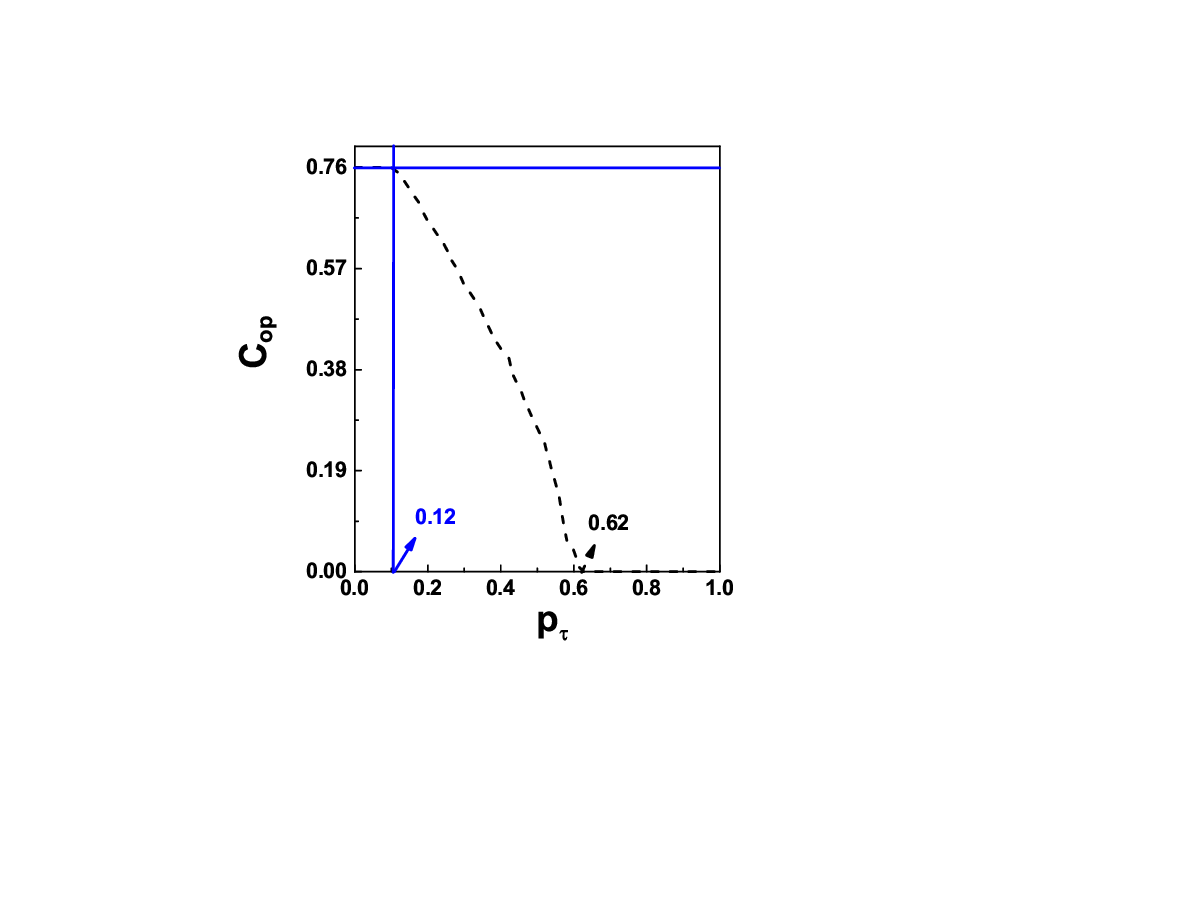}
\caption{Dependence of the optimal initial entanglement $C_{op}$ on the decoherence parameter $p_{\tau}$,
when Alice and Bob's state are in the DPC. The parameters are chosen as $\omega=1, T=3$.}
\end{figure}

We first analyze the effect of the initial entanglement $C$ on QSLT when Alice and Bob's
state are in the DPC. The analysis results are described in Fig. 3. We discover that the influence 
of the initial entanglement $C$ on QSLT depends on the decoherence parameter $p_{\tau}$.
For relatively small values of $p_{\tau}$ (e.g., $p_{\tau}=0.01$ in Fig. 3(a)), the increase of $C$ can 
reduce the QSLT of the system. However, for larger values of $p_{\tau}$ (e.g., $p_{\tau}=0.6,0.8$ in Figs. 3(c) and (d)),
QSLT increases monotonically with the increase of $C$, suggesting that the larger the 
initial entanglement $C$, the weaker the capacity for potential speedup of quantum system.
Moreover, for particular values of the decoherence parameter $p_{\tau}$ in Fig. 3(b), 
QSLT first decreases and then increases as the initial entanglement $C$ increases.
These behaviors illustrate that when the decoherence parameter $p_{\tau}$ is fixed, there is an optimal initial entanglement $C_{op}$ to minimize the value of QSLT, 
thus leading to the maximum evolution speed of the system.

Then to further study the relationship between the optimal initial entanglement $C_{op}$ and decoherence parameters $p_{\tau}$,
the variations of the optimal initial entanglement $C_{op}$ with respect to $p_{\tau}$ is plotted in
Fig. 4. Clearly, when the decoherence parameters are limited to the specfic regions $[0,0.12]$  and $[0.62,1]$, 
the optimal initial entanglement are $C_{max}=0.76$ and $C_{min}=0$, respectively. However, 
it is worth noting that, $C_{op}$ decreases monotonically as $p_{\tau}$ increases, 
when the decoherence parameter $p_{\tau}$ is in a particular region $[0.12,0.62]$.
That is to say, in the DPC, the larger $p_{\tau}$ may 
need the smaller $C$ to excite the maximum speed of evolution of the system.

\subsection{\textbf{In the BFC and BPFC}}

\begin{figure}[tbh]
\includegraphics*[bb=51 167 317 421,width=8cm, clip]{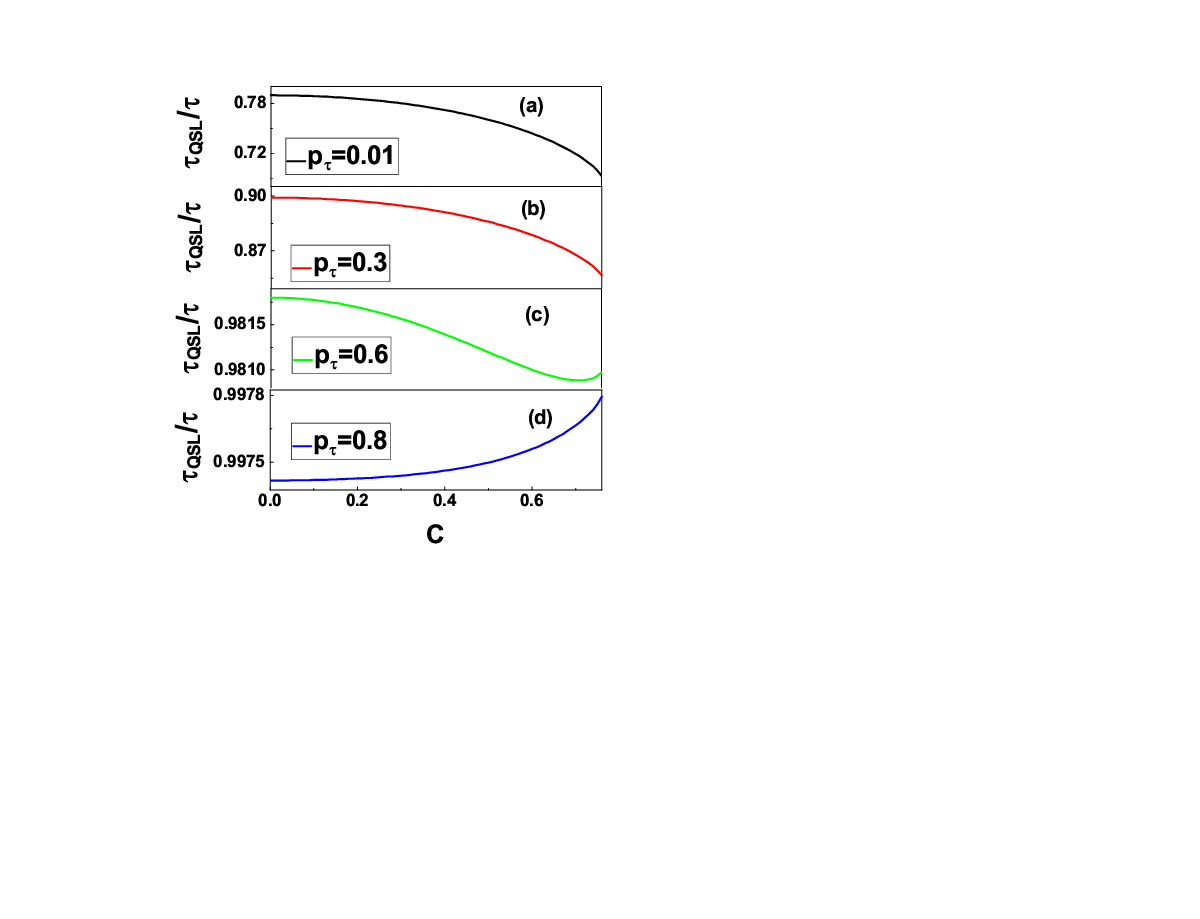}
\caption{(Color online) QSLT for a given three-qubit state,
quantified by $\tau_{QSL}/\tau$ as a function of the initial entanglement
 $C$ with $\omega=1, T=3$, when Alice and Bob's state
are in the BFC.}
\end{figure}

Now we explore the effects of the initial entanglement $C$ on QSLT when  Alice and Bob's
state are in the BFC. Figure 5 shows the results of our analysis for
QSLT as a function of the initial entanglement $C$.
We discover that, by fixing the decoherence parameter $p_{\tau}$, QSLT changes 
monotonically (decreasing or increasing) or non-monotonically (decreasing first and then increasing) 
with the increase of $C$. This implies that, in the BFC,
the maximum evolution speed of the system can be obatained by choosing the optimal initial entanglement $C_{op}$. 
Here $C_{op}$ is closely related to the decoherence parameter $p_{\tau}$, as shown in Fig. 6.
We find that $C_{op}$ decreases monotonically with the increase of $p_{\tau}$ in a specific region $[0.58,0.72]$, 
which is in sharp contrast to $p_{\tau}$ in other regions. These behaviors are analogous to the ones found
before for the case of DPC. A larger (smaller) $p_{\tau}$ and a smaller (larger) $C$ would maximize the capacity 
for potential speedup of quantum system.
Furthermore, since the QSLT expression of the system in the BPFC is the same as that of the BFC, 
one can get the same result in the BPFC as the system in the above channel (i.e., BFC).

\begin{figure}[tbh]
\includegraphics*[bb=56 116 334 387,width=8cm, clip]{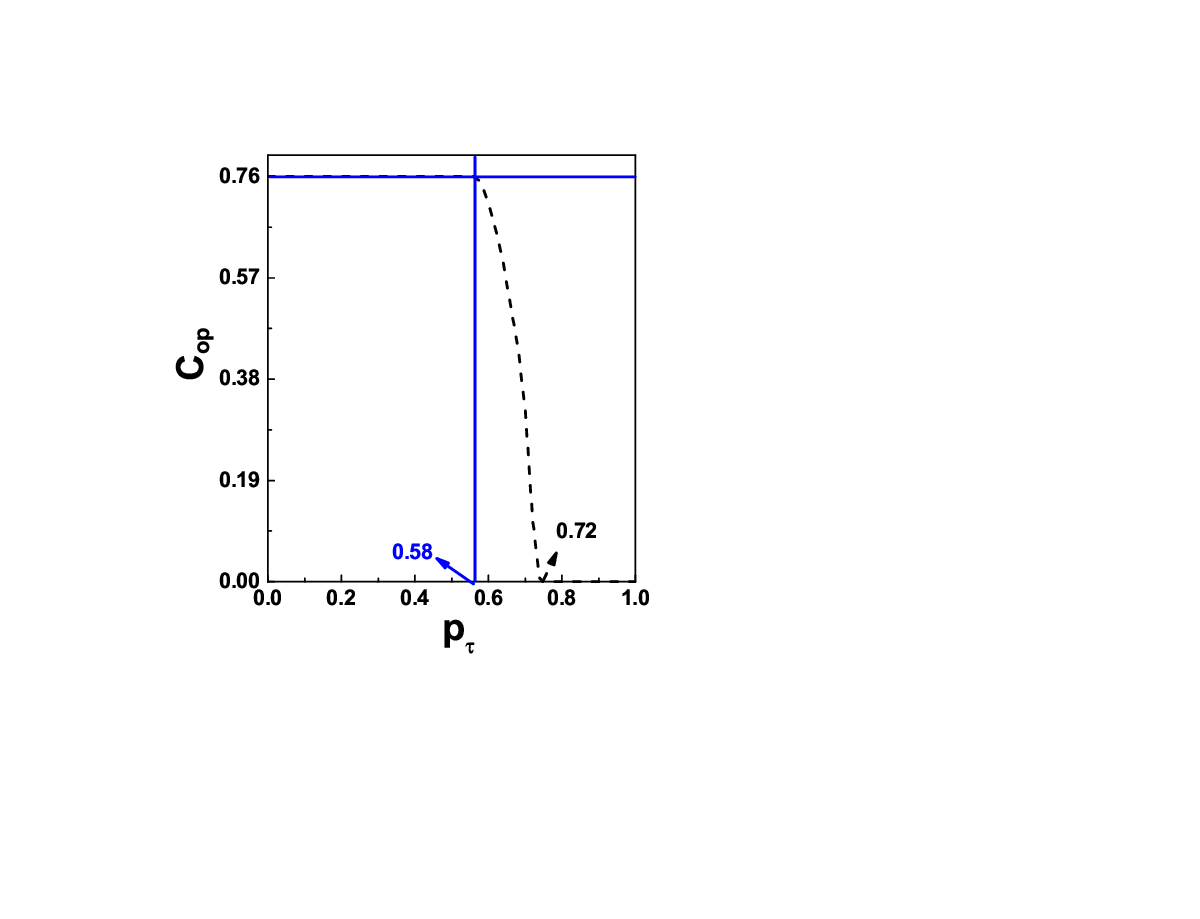}
\caption{Dependence of the optimal initial entanglement $C_{op}$ on the decoherence parameter $p_{\tau}$,
when Alice and Bob's state are in the BFC. The parameters are chosen as $\omega=1, T=3$.}
\end{figure}

\section{\textbf{{Conclusion}}}

In this paper, we have investigated the QSLT of open system  
in the background of a Schwarzschild black hole, as well as four different typical noisy environments, DPC, BFC, BPFC and PFC. 
It is shown that, in some specific Pauli noisy channel, the evolution speed of the system
can be enhanced with the increase of the Hawking temperature.
More specifically, for the initial entangled state in the DPC, BFC, and BPFC,
enhancing the Hawking temperature can speedup the evolution of quantum system.
However, the evolution speed of the system cannot be changed as the Hawking temperature increases, 
revealed by the analytical expression of QSLT when the initial state of the system is unentangled state.
Moreover, by fixing the decoherence parameter in other channels except the PFC, 
we have found that there is an optimal initial entanglement value that 
can make the QSLT of the system reach the minimum, thus leading to the maximum evolution speed of the system.
Now experimental simulations of the Hawking effect rely on the evolution of
quantum states \cite{Horstmannq,MunozdeNova,Steinhauerq}. Therefore, enhancing the evolution speed of the quantum state can decrease the simulation time,
thereby improving the simulation efficiency and robustness under environmental noise.
This implies our results can be beneficial to the experimental simulations of relativistic effects.

\section{\textbf{{Acknowledgements}}}
This work was supported by NSFC under grants Nos. 11574022,
11434015, 61227902, 61835013, 11611530676, KZ201610005011,
the National Key R\&D Program of China under grants Nos. 2016YFA0301500,
SPRPCAS under grants No. XDB01020300, XDB21030300.

\end{document}